\documentclass
[aps,pra,amsfonts,amssymb,twocolumn,amsmath,preprintnumbers,nofootinbib,floatfix,
showpacs,superscriptaddress]{revtex4-1}%
\usepackage[dvips]{graphics}
\usepackage{graphicx}
\usepackage{bm}
\usepackage{amsmath}
\usepackage{amsfonts}
\usepackage{amssymb}%
\providecommand{\U}[1]{\protect\rule{.1in}{.1in}}
\begin{document}

\title{Semiclassical theory of spin-orbit torques in disordered multiband electron systems}
\author{Cong Xiao}
\affiliation{Department of Physics, The University of Texas at Austin, Austin, Texas 78712, USA}

\author{Qian Niu}
\affiliation{Department of Physics, The University of Texas at Austin, Austin, Texas 78712, USA}
\affiliation{ICQM and CICQM, School of Physics, Peking University, Beijing 100871, China}
\begin{abstract}
We study spin-orbit torques (SOT) in non-degenerate multiband electron systems
in the weak disorder limit. In order to have better physical transparency a
semiclassical Boltzmann approach equivalent to the Kubo diagrammatic approach
in the non-crossing approximation is formulated. This semiclassical framework
accounts for the interband-coherence effects induced by both the electric
field and static impurity scattering. Using the two-dimensional Rashba
ferromagnet as a model system, we show that the antidamping-like SOT arising
from disorder-induced interband-coherence effects is very sensitive to the
structure of disorder potential in the internal space and may have the same
sign as the intrinsic SOT in the presence of spin-dependent disorder. While
the cancellation of this SOT and the intrinsic one occurs only in the case of
spin-independent short-range disorder.
\end{abstract}
\maketitle
%\pacs{72.10.-d, 72.10.Bg, 72.25.-b, 72.15.Gd}
%\affiliation{International Center for Quantum Materials, Peking University, Beijing 100871, China}

\section{Introduction}

Disorder effects to nonequilibrium properties of Bloch electrons in solids is
a basic issue in the condensed matter physics. In many instances a relaxation
time approximation is employed to account for the disorder effects
\cite{Ziman1960}. However, this conventional treatment is not enough in some
transport phenomena related to the spin-orbit coupling such as the spin Hall
and anomalous Hall effects \cite{Nagaosa2010,Sinova2015}. To explain these
phenomena, intriguing disorder-induced interband-coherence effects have been
discussed extensively \cite{Nagaosa2010,Sinova2015,Kovalev2010}.

In inversion-asymmetric materials with local magnetization coupled to
conduction electrons in spin-orbit coupled bands, an electric field induces a
nonequilibrium spin-polarization which exerts a torque on the magnetization.
This torque relies on the spin-orbit coupling and is termed spin-orbit torque
(SOT) \cite{Hals2014}. Disorder effects on the SOT have been treated in most
studies by just a constant lifetime approximation
\cite{Zhang2008,Duine2012,Lee2015,Kurebayashi2014,Li,Freimuth2014PRB} or a
single transport relaxation time \cite{Freimuth2016}, leaving the
disorder-induced interband-coherence effects largely unexplored \cite{note}.

The semiclassical Boltzmann transport theory which works in the weak
scattering limit with well-defined multiple-band structure is appealing in its
ability to obtain intuitive pictures \cite{Ziman1960}. Existing semiclassical
Boltzmann theories for SOTs account for the intrinsic contribution
\cite{Duine2012,Lee2015} due to the electric-field-induced interband-coherence
effect \cite{Garate2009,Kurebayashi2014} and a field-like contribution
proportional to the relaxation time or electron lifetime
\cite{Zhang2008,Duine2012,Lee2015}. However, the interband-coherence effects
induced by static impurity scattering cannot be treated by the conventional
Boltzmann equation where the only role of scattering is to equilibrate the
acceleration of electrons by the electric field. Successful inclusion of
disorder-induced interband-coherences into the semiclassical Boltzmann
formalism has recently been realized in the context of the anomalous Hall
effect \cite{Sinitsyn2006,Sinitsyn2007,Sinitsyn2008}, but that formalism
cannot be directly applied to study other spin-related nonequilibrium
phenomena such as the SOT.

In the present paper we focus on SOTs in two-dimensional (2D) Rashba
ferromagnets with the magnetization perpendicular to the 2D plane in the case
of both Rashba bands partially occupied in the weak disorder limit. This
isotropic model enables us to obtain analytical results. We find that the
antidamping-like SOT arising from disorder-induced interband-coherence may
have the same sign as the intrinsic SOT in the presence of spin-dependent
disorder, and the cancellation between them occurs only in the case of
spin-independent short-range (pointlike) disorder. Thus a careful analysis of
different structures of disorder potentials in the internal space is indispensable for the study of SOT.
Moreover, our results imply that, for finite-range or long-range disorder,
other fine details of disorder also need to be carefully accounted for beyond
simple phenomenological treatment.

In order to have better physical transparency, we formulate the analysis in a
semiclassical Boltzmann framework taking into account the interband-coherence
effects due to both the electric field and static impurities in non-degenerate
multiband electron systems. Besides making use of the modified semiclassical
Boltzmann equation \cite{Sinitsyn2007,Sinitsyn2008} developed in the
semiclassical theory of anomalous Hall effect, the scattering-induced
modification to conduction-electron states plays a vital role in this
formalism whose validity is not limited to anomalous Hall effect and SOT.
Regarding the disorder-induced interband-coherence contributions, the
equivalence between the semiclassical theory and microscopic linear response
theory in the weak disorder limit under the non-crossing approximation is established.

The rest of the paper is organized as follows. The semiclassical formulation
is present in Sec. II, whereas model calculations are given in Sec. III. We
make some discussions and conclude the paper in Sec. IV. Appendices A -- C
include some supplementary discussions.

\section{Semiclassical picture}

In the semiclassical version of linear response analysis, the average value of
an observable $A$ (quantum mechanically, Hermitian operator $\hat{A}$, which
can represent a vector, scalar, etc) in the presence of a dc weak uniform
electric field $\mathbf{E}$ and weak static disorder is given by
\cite{Ziman1960}
\begin{equation}
A=\sum_{l}f_{l}A_{l}. \label{semiclassical-1}%
\end{equation}
Here $f_{l}$ is the semiclassical Boltzmann distribution function governed by
the linearized semiclassical Boltzmann equation, $A_{l}$ represents the amount
of $A$ carried by the conduction-electron state denoted by index $l$. In the
present paper we consider non-degenerate multiband electron (hole) systems in
the weak disorder limit, and do not consider thermal related effects. We will
show that, by properly considering $f_{l}$ and $A_{l}$, this semiclassical
framework takes into account the interband-coherence effects induced by both
the electric field and static impurities.

The presence of weak electric field and impurity scattering modifies the
conduction-electron state, making $A_{l}$ deviate from its pure band-value
$A_{l}^{0}\equiv\langle l|\hat{A}\mathbf{|}l\rangle$. Here $|l\rangle$ is the
eigenstate (Bloch state) of disorder-free Hamiltonian $\hat{H}_{0}$. In
equilibrium, $A_{l}$ is modified to $A_{l}=A_{l}^{0}+\delta^{ex}A_{l}$, where
$\delta^{ex}A_{l}$ is related to the scattering-induced correction to Bloch
state $|l\rangle$. Thus the semiclassical expression for the equilibrium value
of $A$ is $A_{0}=\sum_{l}f_{l}^{0}\left(  A_{l}^{0}+\delta^{ex}A_{l}\right)  $
with $f_{l}^{0}$ the Fermi distribution function. Because $\delta^{ex}A_{l}$
is at least linear in the impurity concentration, in the weak disorder limit
one has the conventional expression $A_{0}=\sum_{l}f_{l}^{0}A_{l}^{0}$.
However, in the presence of the electric field, the out-of-equilibrium
distribution function has a component inversely proportional to the impurity
concentration, and thus $\delta^{ex}A_{l}$ contributes to nonequilibrium
phenomena even in the weak disorder limit. Besides, the electric field also
induces a correction $\delta^{in}A_{l}$ to $A_{l}$ related to the so-called
intrinsic contribution \cite{Zhang2005,Lee2015}. As will be explained in Sec.
II. B, in the linear response regime and weak disorder limit, $\delta
^{ex}A_{l}$ and $\delta^{in}A_{l}$ are independent.

In the rest of this section we present formal expressions for $f_{l}$ and
$A_{l}$, and describe how the interband-coherence effects are included into
the semiclassical formalism.

\subsection{Semiclassical distribution function $f_{l}$}

In this subsection we briefly describe the modified semiclassical Boltzmann
equation proposed by Sinitsyn et al. \cite{Sinitsyn2007,Sinitsyn2008} to
determine the distribution function $f_{l}$.

The linearized semiclassical Boltzmann equation for electrons (charge e) in
nonequilibrium steady-states in the presence of elastic electron-impurity
scattering takes the form \cite{Sinitsyn2007}:%
\begin{equation}
e\mathbf{E}\cdot\mathbf{v}_{l}^{0}\frac{\partial f^{0}}{\partial\epsilon_{l}%
}=-\sum_{l^{\prime}}\omega_{l,l^{\prime}}\left(  f_{l}-f_{l^{\prime}}%
-\frac{\partial f^{0}}{\partial\epsilon_{l}}e\mathbf{E}\cdot\delta
\mathbf{r}_{l\prime,l}\right)  . \label{SBE}%
\end{equation}
Here $\mathbf{v}_{l}^{0}$ is the band velocity, $\omega_{l,l^{\prime}}$ is the
semiclassical scattering rate ($l^{\prime}\rightarrow l$) calculated by the
golden rule, $\delta\mathbf{r}_{l^{\prime},l}$\ denotes the coordinate-shift
\cite{Sinitsyn2006} during the scattering and reads $\delta\mathbf{r}%
_{l\prime,l}=\langle u_{l^{\prime}}|i\partial_{\mathbf{k}^{\prime}}%
|u_{l\prime}\rangle-\langle u_{l}|i\partial_{\mathbf{k}}|u_{l}\rangle-\left(
\partial_{\mathbf{k}^{\prime}}+\partial_{\mathbf{k}}\right)  \arg\left(
\langle l^{\prime}|\hat{V}|l\rangle\right)  $ in the lowest nonzero Born
approximation \cite{Sinitsyn2006}. $|l\rangle=|\eta\mathbf{k}\rangle$ is the
Bloch state with eigenenergy $\epsilon_{l}\equiv\epsilon_{\mathbf{k}}^{\eta}$,
$\eta$ is the band index and $\mathbf{k}$ the crystal momentum. $\arg\left(
..\right)  $ denotes the phase of a complex number.

The distribution function is decomposed into \cite{Sinitsyn2007}%
\begin{equation}
f_{l}=f_{l}^{0}+g_{l}^{n}+g_{l}^{a}%
\end{equation}
with $g_{l}^{n}$ equilibrating the acceleration of electrons in the electric
field between scattering events and the anomalous distribution function
$g_{l}^{a}$ describing the effect of electric field working during the
coordinate-shift process. The coordinate-shift is a disorder-induced
interband-coherence effect
\cite{Sinitsyn2006,Sinitsyn2007,Sinitsyn2008,Nagaosa2010} (i.e., related to
interband virtual transitions induced by static impurity scattering) and can
be directly related to the momentum-space Berry curvature \cite{Niu2010} in
some simple cases \cite{Sinitsyn2006,Sinitsyn2007,Sinitsyn2008}. Thus the
anomalous distribution function $g_{l}^{a}$\ is also related to the
disorder-induced interband-coherence.

Under the Gaussian disorder approximation (we restrict to this approximation
throughout this paper), $g_{l}^{n}$ can be further divided into
\cite{Sinitsyn2008,Xiao2017AHE}%
\begin{equation}
g_{l}^{n}=g_{l}^{2s}+g_{l}^{sk-in},
\end{equation}
where $g_{l}^{2s}$ is value of $g_{l}^{n}$ in the lowest Born approximation
($\omega_{l,l^{\prime}}\rightarrow\omega_{l,l^{\prime}}^{2s}$), $g_{l}%
^{sk-in}$ is responsible for the so-called intrinsic-skew-scattering arising
in higher Born orders due to the asymmetry $\omega_{l,l^{\prime}}\neq
\omega_{l^{\prime},l}$ under the Gaussian disorder
\cite{Sinitsyn2008,Xiao2017AHE}. Here we mention that the
intrinsic-skew-scattering is a delicate disorder effects related also to the
interband-coherence (More discussions can be found in Appendix B).

In the presence of pointlike scalar impurities, one can easily verify that
$g_{l}^{sk-in}$ and $g_{l}^{a}$\ do not depend on either the impurity density
or the scattering strength \cite{Xiao2017AHE}, and $g_{l}^{2s}$ is inversely
proportional to the impurity density. A systematic analysis of Eq. (\ref{SBE})
under the non-crossing approximation in isotropic 2D electron systems with
multiple Fermi circles has been presented in Ref. \onlinecite{Xiao2017AHE}.
Anisotropy in band structures or impurity potentials complicates the
analytical treatment, but is not a severe obstacle in numerical solutions
\cite{Freimuth2016}.

\subsection{Scattering and electric-field modified $A_{l}$}

In this subsection we obtain the expression for $A_{l}$ taking into account
the interband-coherences induced by both the electric field and static disorder.

To do this, we firstly deal with the case where there is only the electric
field or only disorder. The electric-field-induced correction to $A_{l}^{0}$
reads $\delta^{in}A_{l}=2\operatorname{Re}\langle l|\hat{A}|\delta
^{\mathbf{E}}l\rangle$, arising from the electric-field-induced
interband-virtual-transition correction%
\[
|\delta^{\mathbf{E}}l\rangle=-i\hbar e\mathbf{E\cdot}\sum_{\eta^{\prime}%
\neq\eta}|\eta^{\prime}\mathbf{k}\rangle\langle u_{\mathbf{k}}^{\eta^{\prime}%
}|\mathbf{\hat{v}}|u_{\mathbf{k}}^{\eta}\rangle/\left(  \epsilon_{\mathbf{k}%
}^{\eta}-\epsilon_{\mathbf{k}}^{\eta^{\prime}}\right)  ^{2}%
\]
to the Bloch state $|l\rangle=|\eta\mathbf{k}\rangle$. Here $|\mathbf{k}%
\rangle$ and $|u_{\mathbf{k}}^{\eta}\rangle$ are the plane-wave and periodic
parts of $|\eta\mathbf{k}\rangle$, respectively. $\mathbf{\hat{v}}$ is the
velocity operator. $\delta^{in}A_{l}$ is an interband-coherence effect induced
solely by the electric field.

Similarly, the scattering-induced correction $\delta^{ex}A_{l}$ stems from
interband-coherence effects in the scattering process. To obtain this part,
one can notice that the Bloch state is also modified by the scattering
according to the Lippmann-Schwinger equation $|l^{s}\rangle=|l\rangle+\left(
\epsilon_{l}-\hat{H}_{0}+i\epsilon\right)  ^{-1}\hat{T}|l\rangle$. Here
$\hat{T}=\hat{V}+\hat{V}\left(  \epsilon_{l}-\hat{H}_{0}+i\epsilon\right)
^{-1}\hat{T}$ is the T-matrix, $\hat{V}$ is the disorder potential. Thus
$\delta^{ex}A_{l}$ is related to $\left\langle 2\operatorname{Re}\langle
l|\hat{A}|\delta^{s}l\rangle+\langle\delta^{s}l|\hat{A}|\delta^{s}%
l\rangle\right\rangle _{c}$, where $|\delta^{s}l\rangle\equiv|l^{s}%
\rangle-|l\rangle$ represents the scattering-induced correction to the Bloch
state and $\left\langle ..\right\rangle _{c}$ denotes the average over
disorder configurations. Here we only consider \cite{note-SOscattering}
disorder potential-free $\hat{A}$. By the Lippmann-Schwinger equation, in the
lowest nonzero order of the disorder potential we get
\[
\left\langle \langle\delta^{s}l|\hat{A}|\delta^{s}l\rangle\right\rangle
_{c}=\sum_{l^{\prime}l^{\prime\prime}}\frac{\left\langle V_{ll\prime
}V_{l\prime\prime l}\right\rangle _{c}\langle l^{\prime}|\hat{A}%
\mathbf{|}l^{\prime\prime}\rangle}{\left(  \epsilon_{l}-\epsilon_{l\prime
}-i\epsilon\right)  \left(  \epsilon_{l}-\epsilon_{l\prime\prime}%
+i\epsilon\right)  }%
\]
and%
\[
\left\langle 2\operatorname{Re}\langle l|\hat{A}|\delta^{s}l\rangle
\right\rangle _{c}=2\operatorname{Re}\sum_{l^{\prime}l^{\prime\prime}}%
\frac{\left\langle V_{l\prime l\prime\prime}V_{l\prime\prime l}\right\rangle
_{c}\langle l|\hat{A}\mathbf{|}l^{\prime}\rangle}{\left(  \epsilon
_{l}-\epsilon_{l\prime}+i\epsilon\right)  \left(  \epsilon_{l}-\epsilon
_{l\prime\prime}+i\epsilon\right)  }.
\]
Both of them contain intraband and interband matrix elements of $\hat{A}$\ in
the band representation. In the weak disorder limit the intraband terms will
be ignored because they are just trivial renormalization effects
\cite{Sinitsyn2007}. Only the interband terms\ are left as nontrivial
corrections to $A_{l}^{0}$ in the weak disorder limit, because they are
interband-coherence effects induced by impurities.

Now we turn to the case where both the electric field and disorder are
present. In equilibrium with disorder, $A_{l}=A_{l}^{0}+\delta^{ex}A_{l}$. The
application of the electric field modifies both $A_{l}^{0}$ and $\delta
^{ex}A_{l}$. However, in the linear response regime, only the
electric-field-induced correction to $A_{l}^{0}$ contributes to nonequilibrium
phenomena in the weak disorder limit and reads $\delta^{in}A_{l}%
=2\operatorname{Re}\langle l|\hat{A}|\delta^{\mathbf{E}}l\rangle$, just the
same as that in the absence of disorder. Therefore we conclude that in the
linear response regime and the weak disorder limit, the effects of electric
field and disorder on $A_{l}$ are independent and thus can be treated
separately. Accordingly, taking into account the electric-field- and
scattering-induced interband-coherence effects, $A_{l}$ can be written as%
\begin{equation}
A_{l}=A_{l}^{0}+\delta^{ex}A_{l}+\delta^{in}A_{l}. \label{semi-A-inter}%
\end{equation}
The intrinsic correction due to the electric-field-induced interband-coherence
is%
\begin{equation}
\delta^{in}A_{l}=\hbar e\sum_{\eta^{\prime}\neq\eta}\frac{2\operatorname{Im}%
\langle\eta\mathbf{k}|\hat{A}|\eta^{\prime}\mathbf{k}\rangle\langle
u_{\mathbf{k}}^{\eta^{\prime}}|\mathbf{\hat{v}\cdot E}|u_{\mathbf{k}}^{\eta
}\rangle}{\left(  \epsilon_{\mathbf{k}}^{\eta}-\epsilon_{\mathbf{k}}%
^{\eta^{\prime}}\right)  ^{2}}, \label{in}%
\end{equation}
whereas the extrinsic correction due to the scattering-induced
interband-coherence reads
\begin{equation}
\delta^{ex}A_{l}=\delta_{1}^{inter}A_{l}+\delta_{2}^{inter}A_{l}, \label{ex}%
\end{equation}
with
\begin{align}
\delta_{1}^{inter}A_{l}  &  =\sum_{\eta^{\prime}\mathbf{k}^{\prime}}\sum
_{\eta^{\prime\prime}\neq\eta^{\prime}}\label{inter-1}\\
\times &  \frac{\left\langle \langle\eta\mathbf{k}|\hat{V}\mathbf{|}%
\eta^{\prime}\mathbf{k}^{\prime}\rangle\langle\eta^{\prime\prime}%
\mathbf{k}^{\prime}|\hat{V}\mathbf{|}\eta\mathbf{k}\rangle\right\rangle
_{c}\langle\eta^{\prime}\mathbf{k}^{\prime}|\hat{A}\mathbf{|}\eta
^{\prime\prime}\mathbf{k}^{\prime}\rangle}{\left(  \epsilon_{\mathbf{k}}%
^{\eta}-\epsilon_{\mathbf{k}^{\prime}}^{\eta^{\prime}}-i\epsilon\right)
\left(  \epsilon_{\mathbf{k}}^{\eta}-\epsilon_{\mathbf{k}^{\prime}}%
^{\eta^{\prime\prime}}+i\epsilon\right)  }\nonumber
\end{align}
and%
\begin{align}
\delta_{2}^{inter}A_{l}  &  =2\operatorname{Re}\sum_{\eta^{\prime}\neq\eta
}\sum_{\eta^{\prime\prime}\mathbf{k}^{\prime\prime}}\label{inter-2}\\
\times &  \frac{\left\langle \langle\eta^{\prime}\mathbf{k}|\hat{V}%
\mathbf{|}\eta^{\prime\prime}\mathbf{k}^{\prime\prime}\rangle\langle
\eta^{\prime\prime}\mathbf{k}^{\prime\prime}|\hat{V}\mathbf{|}\eta
\mathbf{k}\rangle\right\rangle _{c}\langle\eta\mathbf{k}|\hat{A}\mathbf{|}%
\eta^{\prime}\mathbf{k}\rangle}{\left(  \epsilon_{\mathbf{k}}^{\eta}%
-\epsilon_{\mathbf{k}}^{\eta^{\prime}}+i\epsilon\right)  \left(
\epsilon_{\mathbf{k}}^{\eta}-\epsilon_{\mathbf{k}^{\prime\prime}}%
^{\eta^{\prime\prime}}+i\epsilon\right)  }.\nonumber
\end{align}

Under a local phase transformation, $\delta^{ex}A_{l}$\ remains unchanged for
disorder potential \cite{note-SOscattering} $\hat{V}\left(  \mathbf{r}\right)
$. In fact all the three terms of $A_{l}$ in Eq. (\ref{semi-A-inter}) are
gauge invariant and can be regarded as the basic ingredients of a
semiclassical theory.

In the case of $\hat{A}=\mathbf{\hat{v}}$, Eq. (\ref{semi-A-inter}) is just
the velocity of semiclassical electrons appeared in the semiclassical theory
of anomalous Hall effect \cite{Sinitsyn2007,Nagaosa2010}: $\mathbf{v}_{l}^{0}$
is the band velocity, $\delta^{in}\mathbf{v}_{l}$ is the Berry-curvature
anomalous velocity \cite{Niu2010}, and one can show that $\delta
^{ex}\mathbf{v}_{l}$ is consistent with the semiclassical side-jump velocity
$\mathbf{v}_{l}^{sj}$ proposed by Sinitsyn et al.
\cite{Sinitsyn2006,Sinitsyn2007,Sinitsyn2008} as well as Luttinger's quantum
transport theory on the anomalous Hall effect \cite{Luttinger1958} (Detailed
discussions are present in Appendix A). This consistency indicates that the
so-called side-jump velocity can also be understood as arising from
scattering-induced modifications to the Bloch state.

More importantly, this consistency implies that $\delta^{ex}A_{l}$ provides a
generalization of the semiclassical side-jump velocity into physical
quantities besides the electric current (velocity). In the spin Hall effect
where the spin is not conserved in spin-orbit-coupled bands, a semiclassical
Boltzmann analysis of disorder-induced interband-coherences is still absent.
This is partly due to, in our opinion, the lack of a spin-current-counterpart
of the side-jump velocity \cite{note-SHE}. Similarly, the lack of a
SOT-counterpart of the side-jump velocity has impeded the development of
semiclassical Boltzmann theories to SOT. Now the identification of
$\delta^{ex}A_{l}$ provides the counterpart of the side-jump velocity for the
case of physical quantities besides the electric current. Also the
identification of $\delta^{ex}A_{l}$ helps establish the equivalence between
the semiclassical theory on the disorder-induced interband-coherence transport
and diagrammatic perturbation theories in the weak disorder limit under the
non-crossing approximation, as demonstrated in Appendix B.

\subsection{Semiclassical expression of linear response in the weak disorder
limit}

In the linear response regime we get the following semiclassical Boltzmann
expression for $\delta A\equiv A-A_{0}$:
\begin{equation}
\delta A=\sum_{l}\left(  A_{l}^{0}+\delta^{ex}A_{l}\right)  \left(
f_{l}-f_{l}^{0}\right)  +\sum_{l}\left(  \delta^{in}A_{l}\right)  f_{l}^{0}.
\label{semiclassical}%
\end{equation}
The first and second terms on the right hand side (rhs) are extrinsic and
intrinsic \cite{Zhang2005} contributions, respectively.

In the weak disorder limit up to the zeroth order of total disorder
concentration and scattering strength, one has
\begin{align}
\delta A  &  =\sum_{l}A_{l}^{0}g_{l}^{2s}+\sum_{l}A_{l}^{0}\left(  g_{l}%
^{a}+g_{l}^{sk-in}\right) \nonumber\\
&  +\sum_{l}\left(  \delta^{ex}A_{l}\right)  g_{l}^{2s}+\sum_{l}\left(
\delta^{in}A_{l}\right)  f_{l}^{0}. \label{SBE response}%
\end{align}
The first term at the rhs is the conventional Boltzmann result in the lowest
Born order, the second term includes contributions from the anomalous
distribution function and intrinsic-skew-scattering. The last two terms arise
from interband-coherence corrections to the semiclassical value of $A_{l}$ in
Eq. (\ref{semiclassical-1}).

Below we label the terms on the rhs of Eq. (\ref{SBE response}) as
$\delta^{2s}A=\sum_{l}A_{l}^{0}g_{l}^{2s}$, $\delta^{sj}A=\sum_{l}\left(
\delta^{ex}A_{l}\right)  g_{l}^{2s}$, $\delta^{adis}A=\sum_{l}A_{l}^{0}%
g_{l}^{a}$, $\delta^{sk-in}A=\sum_{l}A_{l}^{0}g_{l}^{sk-in}$ and $\delta
^{in}A=\sum_{l}\left(  \delta^{in}A_{l}\right)  f_{l}^{0}$. As stated in the
past two subsections, disorder-induced interband-coherence effects are
included in $\delta^{ex}A_{l}$, $g_{l}^{a}$ and $g_{l}^{sk-in}$, the
disorder-induced interband-coherence contribution (labeled by $\delta^{SJ}A$)
to $\delta A$\ is thus
\begin{equation}
\delta^{SJ}A=\delta^{adis}A+\delta^{sk-in}A+\delta^{sj}A. \label{SJ}%
\end{equation}
In the presence of pointlike scalar impurities, all the three terms are
independent of both the impurity density and scattering strength
\cite{Xiao2017AHE}.

In the semiclassical theory of the anomalous Hall effect formulated recently
by Sinitsyn et al. \cite{Sinitsyn2008}, the disorder-induced
interband-coherence contribution (called side-jump effect in that context
\cite{Kovalev2010,Nagaosa2010}) comprises three ingredients: a side-jump
velocity $\mathbf{v}_{l}^{sj}$, the anomalous distribution function $g_{l}%
^{a}$ and intrinsic-skew-scattering $g_{l}^{sk-in}$. As our $\delta
^{ex}\mathbf{v}_{l}$ coincides with $\mathbf{v}_{l}^{sj}$, when applied to the
anomalous Hall effect the present semiclassical formalism is consistent with
that by Sinitsyn et al.

Furthermore, we establish (see Appendix B) an one-to-one correspondence
between the three semiclassical terms at the rhs of Eq. (\ref{SJ}) and special
sets of Feynman diagrams representing the disorder-induced interband-coherence
transport contributions in the band-eigenstate basis under the non-crossing
approximation in the weak disorder limit. This also confirms the validity of
our semiclassical framework.

Equation (\ref{SBE response}) can then be casted into
\begin{equation}
\delta A=\delta^{2s}A+\delta^{SJ}A+\delta^{in}A.
\end{equation}
Here we can mention that, in the weak scattering limit the widely-used
classification of SOT into interband and intraband\ parts does not take into
account $\delta^{SJ}A$, i.e., the disorder-induced interband-coherence effects
far beyond the relaxation time approximation \cite{Chadova2015Separation}.

\section{Model Calculation}

We consider the case where the SOT is related to the nonequilibrium
conduction-electron spin-polarization $\delta\mathbf{S}$ which is coupled to
the local magnetization via the s-d exchange coupling. In the simplified
treatment adopted here, one only calculates $\delta\mathbf{S}$ in the presence
of the driven electric field and disorder
\cite{Zhang2008,Duine2012,Kurebayashi2014,Lee2015,Li}. In this section we
focus on SOTs in 2D Rashba ferromagnets with the magnetization perpendicular
to the 2D plane. In Appendix C we also analyze the case of in-plane
magnetization and scalar pointlike impurities.

The 2D model Hamiltonian is $\hat{H}=\hat{H}_{0}+\hat{V}\left(  \mathbf{r}%
\right)  $, where%
\begin{equation}
\hat{H}_{0}=\frac{\mathbf{\hat{p}}^{2}}{2m}+\frac{\alpha_{R}}{\hbar
}\mathbf{\hat{\sigma}}\cdot\left(  \mathbf{\hat{p}}\times\mathbf{\hat{z}%
}\right)  -J_{ex}\mathbf{\hat{\sigma}\cdot\hat{M}}. \label{model}%
\end{equation}
Here $m$ is the in-plane effective mass of conduction electron, $\mathbf{\hat
{p}=\hbar\hat{k}}$ the 2D momentum, $\mathbf{\hat{\sigma}}=\left(  \hat
{\sigma}_{x},\hat{\sigma}_{y},\hat{\sigma}_{z}\right)  $ are the Pauli
matrices, $\alpha_{R}$ is the Rashba coefficient, $J_{ex}$ the exchange
coupling. $\mathbf{\hat{M}}$ is the direction of the local magnetization and
chosen to be $\mathbf{\hat{M}=\hat{z}}$ for the in-plane isotropic model.
$|u_{\mathbf{k}}^{\eta}\rangle=\frac{1}{\sqrt{2}}\left[  \sqrt{1-\eta
\cos\theta},-i\eta\exp\left(  i\phi\right)  \sqrt{1+\eta\cos\theta}\right]
^{T}$ is the inner eigenstate, where $\eta=\pm$, $\tan\phi=\frac{k_{y}}{k_{x}%
}$, $\cos\theta=J_{ex}/\Delta_{k}$, $\Delta_{k}=\sqrt{\alpha^{2}k^{2}%
+J_{ex}^{2}}$. We only consider the case $\epsilon_{F}>J_{ex}$\textbf{,} i.e.,
both Rashba bands partially occupied. For any energy $\epsilon>J_{ex}$ there
are two iso-energy rings corresponding to the two bands: $k_{\eta}^{2}\left(
\epsilon\right)  =\frac{2m}{\hbar^{2}}\left(  \epsilon-\eta\Delta_{\eta
}\left(  \epsilon\right)  \right)  $ where $\Delta_{\eta}\left(
\epsilon\right)  \equiv\Delta_{k_{\eta}\left(  \epsilon\right)  }%
=\sqrt{\epsilon_{R}^{2}+J_{ex}^{2}+2\epsilon_{R}\epsilon}-\eta\epsilon_{R}$
and $\epsilon_{R}=m\left(  \frac{\alpha_{R}}{\hbar}\right)  ^{2}$. The density
of states in $\eta$ band is $D_{\eta}\left(  \epsilon\right)  =D_{0}%
\frac{\Delta_{\eta}\left(  \epsilon\right)  }{\Delta_{\eta}\left(
\epsilon\right)  +\eta\epsilon_{R}}$ with $D_{0}=\frac{m}{2\pi\hbar^{2}}$.

Hereafter the electric field is applied in the y direction. The intrinsic
nonequilibrium spin-polarization reads%
\begin{equation}
\delta^{in}\mathbf{S=}\sum_{l}\left(  \delta^{in}\mathbf{S}_{l}\right)
f_{l}^{0}=-eE_{y}\frac{\hbar}{2}\frac{J_{ex}\alpha_{R}D_{0}}{J_{ex}%
^{2}+2\epsilon_{R}\epsilon_{F}}\mathbf{\hat{y},}%
\end{equation}
which is parallel to the electric field and contributes an intrinsic
antidamping-like SOT.

It was found in the context of the anomalous Hall effect that the structure of
short-range disorder potential (do not consider spin-orbit scattering) in the
internal space (internal degrees of freedom such as spin and valley) strongly
affects the disorder-induced interband-coherence response \cite{Yang2011}. For
the in-plane isotropic Rashba model where $\mathbf{\hat{S}=}\frac{\hbar}%
{2}\mathbf{\hat{\sigma}}$, according to the structure of the disorder
potential in the $2\times2$ internal space, the pointlike disorder
\cite{note-SOscattering,note-SOscattering-1} can be classified following the
recipe of Yang et al. \cite{Yang2011} as: class A $\hat{V}=V_{A}\hat{\sigma
}_{0}$, class B $\hat{V}=V_{B}\hat{\sigma}_{z}$ and class C $\hat{V}=V_{c}%
\hat{\sigma}_{\pm}/\sqrt{2}$. Here $\hat{\sigma}_{\pm}=$ $\hat{\sigma}_{x}\pm
i\hat{\sigma}_{y}$, $\hat{\sigma}_{0}$ is the $2\times2$ identity matrix.
Details about the theoretical consideration on this classification in in-plane
isotropic systems with $2\times2$ internal space and the realizations of these
scattering classes in practice have been given in Sec. II and Sec. IV of Ref.
\onlinecite{Yang2011}, respectively. It was shown that the disorder-induced
interband-coherence contribution to the anomalous Hall effect in in-plane
isotropic systems with $2\times2$ internal space due to class A disorder is
quite different from that due to classes B and C disorder, even with opposite
signs \cite{Yang2011,Ren2008}. While contributions due to class B and C
disorder are qualitatively similar \cite{Yang2011,note-AHE}. Thus we only take
into account class A and class B disorder to calculate the SOT.

\subsection{Class A disorder}

According to Eqs. (\ref{ex},\ref{inter-1},\ref{inter-2}), we obtain
$\delta_{2}^{inter}\mathbf{S}_{l}=0$ and
\begin{equation}
\delta^{ex}\mathbf{S}_{l}=\delta_{1}^{inter}\mathbf{S}_{l}=-\frac{\hbar}%
{2}\frac{\hbar}{\tau}\frac{\eta J_{ex}}{J_{ex}^{2}+2\epsilon_{R}\epsilon}%
\frac{\alpha_{R}\mathbf{k}_{\eta}\left(  \epsilon\right)  }{2\Delta_{\eta
}\left(  \epsilon\right)  }. \label{sj A}%
\end{equation}
\ $\delta^{ex}\mathbf{S}_{l}$ contributes a nonequilibrium spin-polarization
\begin{equation}
\delta^{sj}\mathbf{S}=\sum_{l}g_{l}^{2s}\delta^{ex}\mathbf{S}_{l}=eE_{y}%
\frac{\hbar}{2}\frac{\alpha_{R}J_{ex}D_{0}}{J_{ex}^{2}+2\epsilon_{R}%
\epsilon_{F}}\mathbf{\hat{y}},
\end{equation}
which completely cancels the intrinsic contribution. For class A disorder
$g_{l}^{2s}=\left(  -\partial_{\epsilon}f^{0}\right)  e\mathbf{E}\cdot
\frac{\hbar\mathbf{k}_{\eta}\left(  \epsilon\right)  }{m}\tau$ has been
obtained before \cite{Xiao2017AHE}, with $\tau=\left(  2\pi n_{im}^{A}%
V_{A}^{2}D_{0}/\hbar\right)  ^{-1}$ the lifetime of Rashba electron with
$n_{im}^{A}$ the density of class A disorder. Moreover, Ref.
\onlinecite{Xiao2017AHE} has shown that the anomalous distribution and the
distribution function for the intrinsic skew scattering cancels each other:
$g_{l}^{a}+g_{l}^{sk-in}=0$. Thus $\delta^{adis}\mathbf{S+}$ $\delta
^{sk-in}\mathbf{S=0}$ and the disorder-induced interband-coherence
contribution is just $\delta^{SJ}\mathbf{S}=\delta^{sj}\mathbf{S}$, then the
total (electric-field-induced plus disorder-induced) interband-coherence
contribution to the nonequilibrium spin-polarization vanishes%
\begin{equation}
\delta^{SJ}\mathbf{S}+\delta^{in}\mathbf{S=0}.
\end{equation}
Thereby $\delta\mathbf{S}=\delta^{2s}\mathbf{S=}\sum_{l}g_{l}^{2s}%
\mathbf{S}_{l}^{0}=-e\alpha_{R}D_{0}\tau E_{y}\mathbf{\hat{x}}$, which is
magnetization-independent and coincides with the well-known Edelstein result
in the nonmagnetic 2D Rashba model with class A disorder \cite{Edelstein}.
This $\delta\mathbf{S}$ perpendicular to both the magnetization and electric
field contributes a field-like SOT.

\subsection{Class B disorder}

In this case, the electron lifetime is still independent of energy and band,
and is given by $\tau=\left(  \sum_{l^{\prime}}\omega_{l^{\prime},l}%
^{2s}\right)  ^{-1}=\left(  \frac{2\pi}{\hbar}n_{im}^{B}V_{B}^{2}D_{0}\right)
^{-1}$ with $n_{im}^{B}$ the density of class B disorder.

According to Eqs. (\ref{ex},\ref{inter-1},\ref{inter-2}), we get $\delta
_{2}^{inter}\mathbf{S}_{l}=0$ and
\begin{equation}
\delta^{ex}\mathbf{S}_{l}=\delta_{1}^{inter}\mathbf{S}_{l}=\frac{\hbar}%
{2}\frac{\hbar}{\tau}\frac{\eta J_{ex}}{J_{ex}^{2}+2\epsilon_{R}\epsilon}%
\frac{\alpha_{R}\mathbf{k}_{\eta}\left(  \epsilon\right)  }{2\Delta_{\eta
}\left(  \epsilon\right)  }. \label{sj B}%
\end{equation}
We note that, for the same $\mathbf{k}_{\eta}\left(  \epsilon\right)  $ the
sign of$\ \delta^{ex}\mathbf{S}_{l}$ is opposite to that in the case of class
A disorder.

The Boltzmann equation (\ref{SBE}) is solved following the recipe given by
Ref. \onlinecite{Xiao2017AHE}. After lengthy calculations we get
\begin{gather}
g_{\eta}^{2s}\left(  \epsilon\right)  =\left(  -\partial_{\epsilon}%
f^{0}\right)  e\mathbf{E}\cdot\frac{\hbar\mathbf{k}_{\eta}\left(
\epsilon\right)  }{m}\tau\frac{\Delta_{-\eta}^{2}\left(  \epsilon\right)
+\epsilon_{R}\epsilon}{J_{ex}^{2}+3\epsilon_{R}\epsilon},\nonumber\\
g_{\eta}^{a}\left(  \epsilon\right)  =\left(  -\partial_{\epsilon}%
f^{0}\right)  \left(  \mathbf{\hat{z}}\times e\mathbf{E}\right)
\cdot\mathbf{k}_{\eta}\left(  \epsilon\right)  \frac{\eta J_{ex}\alpha_{R}%
^{2}}{2\Delta_{\eta}\left(  \epsilon\right)  \left(  J_{ex}^{2}+3\epsilon
_{R}\epsilon\right)  },\nonumber\\
g_{\eta}^{sk-in}\left(  \epsilon\right)  =\frac{J_{ex}^{2}+\epsilon
_{R}\epsilon}{J_{ex}^{2}+3\epsilon_{R}\epsilon}g_{\eta}^{adis}\left(
\epsilon\right)  .
\end{gather}
Then the disorder-induced interband-coherence contributions to the
nonequilibrium spin-polarization in Eq. (\ref{SJ}) are given by%
\begin{align}
\delta^{sj}\mathbf{S}  &  =\sum_{l}g_{l}^{2s}\delta^{ex}\mathbf{S}_{l}%
=\frac{J_{ex}^{2}+\epsilon_{R}\epsilon_{F}}{J_{ex}^{2}+3\epsilon_{R}%
\epsilon_{F}}\delta^{in}\mathbf{S,}\\
\delta^{adis}\mathbf{S}  &  =\sum_{l}g_{l}^{a}\mathbf{S}_{l}^{0}%
=-\frac{\epsilon_{R}\epsilon_{F}}{J_{ex}^{2}+3\epsilon_{R}\epsilon_{F}}%
\delta^{in}\mathbf{S,}%
\end{align}
and%
\begin{equation}
\delta^{sk-in}\mathbf{S}=\sum_{l}g_{l}^{sk-in}\mathbf{S}_{l}^{0}=\frac
{J_{ex}^{2}+\epsilon_{R}\epsilon_{F}}{J_{ex}^{2}+3\epsilon_{R}\epsilon_{F}%
}\delta^{adis}\mathbf{S}.
\end{equation}
Thus the total disorder-induced interband-coherence contribution reads
\begin{equation}
\delta^{SJ}\mathbf{S}=\left[  2\left(  \frac{J_{ex}^{2}+2\epsilon_{R}%
\epsilon_{F}}{J_{ex}^{2}+3\epsilon_{R}\epsilon_{F}}\right)  ^{2}-1\right]
\delta^{in}\mathbf{S}.
\end{equation}
In the large exchange-coupling limit $J_{ex}\gg\sqrt{\epsilon_{R}\epsilon_{F}%
}$ one has $\delta^{SJ}\mathbf{S}\simeq\delta^{in}\mathbf{S}$, the
disorder-induced interband-coherence contribution approximately doubles the
intrinsic nonequilibrium spin-polarization and the corresponding
antidamping-like SOT. While in the opposite limit $J_{ex}\ll\sqrt{\epsilon
_{R}\epsilon_{F}}$, $\delta^{SJ}\mathbf{S}\simeq-\frac{1}{9}\delta
^{in}\mathbf{S}$ and the contribution from disorder-induced
interband-coherences partly cancels the intrinsic nonequilibrium
spin-polarization. In particular, $\delta^{SJ}\mathbf{S}=0$ when $J_{ex}%
^{2}=\left(  \sqrt{2}-1\right)  \epsilon_{R}\epsilon_{F}$.

The total interband-coherence contribution to the nonequilibrium
spin-polarization reads%
\begin{equation}
\delta^{in}\mathbf{S}+\delta^{SJ}\mathbf{S}=-\hbar e\alpha_{R}D_{0}%
\frac{J_{ex}\left(  J_{ex}^{2}+2\epsilon_{R}\epsilon_{F}\right)  }{\left(
J_{ex}^{2}+3\epsilon_{R}\epsilon_{F}\right)  ^{2}}E_{y}\mathbf{\hat{y},}
\label{SOT-yy B Boltzmann}%
\end{equation}
which exerts an antidamping-like torque on the magnetization. Besides,
$\delta^{2s}\mathbf{S}=-e\alpha_{R}D_{0}\tau\frac{J_{ex}^{2}+\epsilon
_{R}\epsilon_{F}}{J_{ex}^{2}+3\epsilon_{R}\epsilon_{F}}E_{y}\mathbf{\hat{x}}$
leads to a field-like torque proportional to $\tau$.

\subsection{Competition between classes A and B}

When the dominant scattering class is tuned (by doping or by varying
temperature \cite{Yang2011}), rich behaviors of SOT are expected even in the
weak disorder limit. In the presence of both class A and class B impurities,
we assume \cite{Yang2011} $\left\langle V_{A}V_{B}\right\rangle _{c}=0$, and
only the main results are given in this case.

Due to $\sum_{\eta}\frac{D_{\eta}}{\Delta_{\eta}}=0$, the electron lifetime is
given by $\tau=\tau_{A}/\left(  1+\zeta\right)  =\left(  \tau_{A}^{-1}%
+\tau_{B}^{-1}\right)  ^{-1}$, where $1/\tau_{A\left(  B\right)  }=2\pi
n_{im}^{A\left(  B\right)  }V_{A\left(  B\right)  }^{2}D_{0}/\hbar$,
$\zeta=\tau_{A}/\tau_{B}$.

In this subsection we write $\delta S_{\alpha}=$ $\chi_{\alpha\beta}E_{\beta}%
$, with $\alpha,\beta=x,y$. Lengthy calculations lead to
\begin{align}
\chi_{xy}  &  =-e\alpha_{R}D_{0}\tau\frac{1-I_{1}}{1-\frac{1-\zeta}{1+\zeta
}I_{1}},\nonumber\\
\chi_{yy}  &  =-\hbar e\alpha_{R}D_{0}\frac{J_{ex}}{J_{ex}^{2}+2\epsilon
_{R}\epsilon_{F}}\frac{\frac{\zeta}{1+\zeta}}{\left(  1-\frac{1-\zeta}%
{1+\zeta}I_{1}\right)  ^{2}},
\end{align}
where $I_{1}=\frac{\epsilon_{R}\epsilon_{F}}{J_{ex}^{2}+2\epsilon_{R}%
\epsilon_{F}}$. \begin{figure}[ptbh]
%Requires \usepackage{graphicx}
\includegraphics[width=0.35\textwidth]{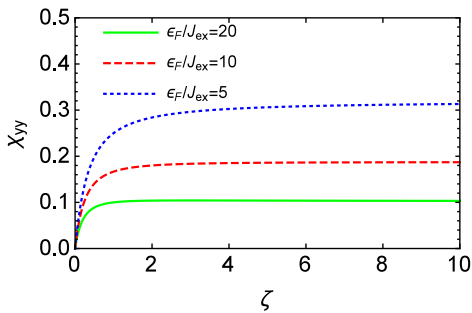} \caption{$\chi_{yy}$
versus $\zeta=\tau_{A}/\tau_{B}$ for fixed values of $\epsilon_{F}/J_{ex}$.
$\chi_{yy}$ is measured in units of $-e\alpha_{R}D_{0}\hbar/J_{ex}$. The plot
shows the crossover from the class A dominated regime to the class B dominated
regime as $\zeta$ increases. Here we set $\epsilon_{R}/J_{ex}=0.1$.}%
\label{fig2}%
\end{figure}\begin{figure}[ptbh]
%Requires \usepackage{graphicx}
\includegraphics[width=0.35\textwidth]{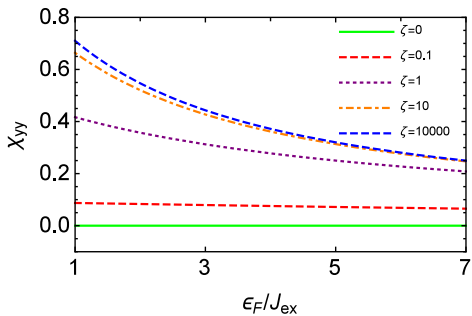} \caption{$\chi
_{yy}$ versus $\epsilon_{F}$ for fixed values of $\zeta$. $\chi_{yy}$ is
measured in units of $-e\alpha_{R}D_{0}\hbar/J_{ex}$, and $\epsilon_{F}$ is
measured in units of $J_{ex}$. We have chosen $\epsilon_{R}/J_{ex}=0.1$ in
plotting the curves.}%
\label{fig3}%
\end{figure}

One can observe that in the limit $\zeta\rightarrow0$ or $\zeta\rightarrow
\infty$, our previous results in Sec. III. A and Sec. III. B are recovered,
and the values of nonequilibrium spin-polarizations (and thus SOTs) vary
continuously as $\zeta$ changes between these two limits. In Fig. 1 we plot
$\chi_{yy}$ as a function of $\zeta$ for fixed values of $\epsilon_{F}$. One
can see that $\chi_{yy}$ increases monotonically as $\zeta$ increases from the
class A dominated regime to the class B dominated regime. The class B
dominated regime is reached at smaller $\zeta$\ for larger $\epsilon_{F}$. In
Fig. 2 we plot $\chi_{yy}$ as a function of $\epsilon_{F}$ for different
values of $\zeta$. As $\zeta$ increases from zero, the curve of $\chi_{yy}$ is
shifted upward from the class A dominated regime due to the increasing
contribution from class B scattering. For the chosen parameter $\epsilon
_{R}/J_{ex}=0.1$, we find that $\zeta=10$ is quite approaching the class B
dominated case $\zeta=10000$. This is consistent with the trend shown in Fig. 1.

\section{Discussion and conclusion}

\subsection{Comparison to other theories}

In the context of the anomalous Hall effect, it has been realized
\cite{Sinitsyn2006PRL,Sinitsyn2007,Nunner2007,Xiao2017AHE} that, in the weak
disorder limit under the non-crossing approximation the semiclassically
obtained disorder-induced interband-coherence contribution (side-jump) is
equivalent to the ladder vertex correction to the bare bubble representing the
intrinsic contribution \cite{note-intrinsic} in the non-chiral basis (just the
spin-$\sigma_{z}$ basis for two-band models such as the Rashba model
\cite{Nunner2007,Xiao2017AHE} and Dirac model
\cite{Sinitsyn2006PRL,Sinitsyn2007}, while the chiral basis means the
band-eigenstate basis \cite{Sinitsyn2007,Kovalev2010}) in Kubo diagrammatic
theories. Only few papers addressed vertex corrections to the intrinsic SOT
\cite{Titov2015,Titov2017,Kohno2014,Ebert2016}, and these calculations do not
give pictures of interband-coherence effects due to employment of the
non-chiral basis.

Regarding the present model, the cancellation between the intrinsic and
disorder-induced interband-coherence contributions in the case of scalar
short-range disorder is consistent with that obtained by calculating the
vertex correction in quantum transport theories \cite{Titov2015,Titov2017}.
For the case of class B disorder, we have also performed a Kubo diagrammatic
calculation \cite{paper} under the non-crossing approximation and obtained the
same weak-disorder-limit result for the SOT as that of the present
semiclassical theory.

\subsection{Relative magnitude of antidamping-like and field-like SOTs}

In the Rashba system with both bands partially occupied, under the good-metal
condition ($\epsilon_{F}\tau/\hbar\gg1$) there are still two different limits
often discussed in literatures. One is the weak disorder limit where the
disorder broadening is much smaller than the band splitting due to Rashba and
exchange couplings \cite{Zhang2008,Lee2015,Li}, the other is the opposite
limit -- diffusive limit \cite{Manchon2012}. If the Rashba and exchange
couplings are both weak, the system may be near the diffusive limit, where the
Boltzmann theory does not work.
%In this limit there does not exist disorder-independent AHE or antidamping-like SOT, thus above definition of intrinsic and side-jump contributions cannot be directly applied. However, one may \textquotedblleft analytically continue\textquotedblright\ the definition of intrinsic and side-jump contributions out of the Boltzmann limit \cite{note-continuation}.

In the weak disorder limit, the antidamping-like SOT from the intrinsic and
disorder-induced interband-coherence contributions is smaller than the
field-like one. However, unlike the longitudinal conductivity whose leading
contribution under the good-metal condition is always proportional to
$\epsilon_{F}\tau/\hbar$, the field-like SOT (proportional to $\chi_{xy}$) is
not proportional to $\epsilon_{F}$ even in the weak disorder limit. Thus as
the system evolves from the weak disorder limit to the diffusive limit, while
the longitudinal electric conductivity remains large, the field-like SOT may
become much smaller and may not remain dominant over the antidamping-like one.
This attracting possibility will be investigated in a separate paper.

\subsection{Neglected contributions}

Very recently, the diagrammatic calculation of the anomalous Hall effect under
the Gaussian disorder beyond the non-crossing approximation has been 
highlighted \cite{Ado2015}. The resulting additional contribution is also
independent of both disorder density and scattering strength in the case of
scalar pointlike impurities in the weak disorder limit. There should also be
corresponding additional contribution to the SOT. As shown by Luttinger nearly 
sixty years ago \cite{Luttinger1958}, this contribution can also be included 
into the semiclassical Boltzmann theory. The concrete calculation is left for future discussion.

We assumed Gaussian disorder as in Refs.
\onlinecite{Kovalev2010,Titov2015,Titov2017,Kohno2014}. Non-Gaussian disorder
is not included in this paper. In the context of the anomalous Hall effect,
non-Gaussian disorder leads to skew scattering contributions which depend on
the scattering time \cite{Nagaosa2010}. In the field of the SOT, the effects
of non-Gaussian disorder can be calculated by the same method as that applied
to the anomalous Hall effect \cite{Nagaosa2010}.

\subsection{Summary}

In summary, we have studied spin-orbit torques in non-degenerate multiband
electron systems by formulating a semiclassical Boltzmann framework in the
weak disorder limit. This semiclassical formulation accounts for
interband-coherence effects induced by both the electric field and static
impurity scattering, and is equivalent to the Kubo diagrammatic approach under
the non-crossing approximation in the weak disorder limit. Using the 2D Rashba
ferromagnets as an example, we showed that the disorder-induced
interband-coherence effects contribute an antidamping-like torque, which is
very sensitive to the structure of disorder potential in the internal space
(spin space for the considered model) and may have the same sign as the
intrinsic spin-orbit torque in the presence of spin-dependent disorder.

We expect these findings are helpful also in understanding spin-orbit torques
in the 2D anti-ferromagnetic Rashba model \cite{Zelezny2017}.
%as well as in comparing the torques in ferromagnetic and anti-ferromagnetic Rashba systems.
The semiclassical framework proposed in this paper can be employed to treat
other nonequilibrium phenomena related to disorder-induced interband-coherence
effects.
%such as the spin Hall effect when the spin is not conserved due to band-structure spin-orbit coupling \cite{note-SHE}.

\begin{acknowledgments}
We acknowledge useful discussions with H. Chen and L. Dong. This work is supported by NBRPC (Grant No. 2013CB921900), DOE (DE-FG03-02ER45958, Division of Materials Science and Engineering), NSF (EFMA-1641101) and Welch Foundation (F-1255). The model calculation in Sec. III is supported by the DOE grant.
\end{acknowledgments}

\appendix

\begin{widetext}
\section{Consistency of our formulas and the side-jump velocity}
In the well-established semiclassical Boltzmann theory of anomalous Hall
effect \cite{Sinitsyn2006,Sinitsyn2007} the side-jump velocity is obtained by
linking it to the coordinate-shift $\mathbf{v}_{l}^{sj}=\sum_{l^{\prime}%
}\omega_{l^{\prime},l}^{2s}\delta\mathbf{r}_{l\prime,l}$. Here we prove that
our $\delta^{ex}\mathbf{v}_{l}=\delta_{1}^{inter}\mathbf{v}_{l}+\delta
_{2}^{inter}\mathbf{v}_{l}$ is consistent with this $\mathbf{v}_{l}^{sj}$. Due
to \cite{note-SOscattering} $\mathbf{\hat{v}=}\frac{1}{i\hbar}\left[
\mathbf{\hat{r},}\hat{H}_{0}\right]  $, we have
\begin{equation}
\delta_{1}^{inter}\mathbf{v}_{l}=\sum_{l^{\prime},l^{\prime\prime}\neq
l^{\prime}}\frac{1}{i\hbar}\left\langle \frac{V_{ll^{\prime}}\langle
l^{\prime}|\mathbf{\hat{r}|}l^{\prime\prime}\rangle V_{l^{\prime\prime}l}%
}{\epsilon_{l}-\epsilon_{l^{\prime}}-i\delta}\frac{\epsilon_{l^{\prime\prime}%
}-\epsilon_{l^{\prime}}}{\epsilon_{l}-\epsilon_{l^{\prime\prime}}+i\delta
}\right\rangle _{c}=2\operatorname{Re}\sum_{l^{\prime},l^{\prime\prime}\neq
l^{\prime}}\frac{i}{\hbar}\left\langle V_{ll^{\prime}}\frac{\langle l^{\prime
}|\mathbf{\hat{r}|}l^{\prime\prime}\rangle V_{l^{\prime\prime}l}}{\epsilon
_{l}-\epsilon_{l^{\prime}}-i\delta}\right\rangle _{c}%
\end{equation}
and%
\begin{align}
\delta_{2}^{inter}\mathbf{v}_{l} &  =\operatorname{Re}\sum_{l^{\prime}\neq
l,l^{\prime\prime}}\frac{2}{i\hbar}\left\langle \frac{\epsilon_{l^{\prime}%
}-\epsilon_{l}}{\epsilon_{l}-\epsilon_{l^{\prime}}+i\delta}\frac{\langle
l|\mathbf{\hat{r}|}l^{\prime}\rangle V_{l^{\prime}l^{\prime\prime}%
}V_{l^{\prime\prime}l}}{\epsilon_{l}-\epsilon_{l^{\prime\prime}}+i\delta
}\right\rangle _{c}\nonumber\\
&  =2\operatorname{Re}\sum_{l^{\prime}\neq l,l^{\prime\prime}}\frac{i}{\hbar
}\left\langle \langle l|\mathbf{\hat{r}|}l^{\prime}\rangle\frac{V_{l^{\prime
}l^{\prime\prime}}V_{l^{\prime\prime}l}}{\epsilon_{l}-\epsilon_{l^{\prime
\prime}}+i\delta}\right\rangle _{c}=2\operatorname{Re}\sum_{l^{\prime
},l^{\prime\prime}\neq l}\frac{-i}{\hbar}\left\langle \langle l^{\prime\prime
}|\mathbf{\hat{r}|}l\rangle\frac{V_{l^{\prime}l^{\prime\prime}}V_{ll^{\prime}%
}}{\epsilon_{l}-\epsilon_{l^{\prime}}-i\delta}\right\rangle _{c},
\end{align}
thus%
\begin{equation}
\delta^{ex}\mathbf{v}_{l}=2\operatorname{Re}\sum_{l^{\prime},l^{\prime\prime
}\neq l^{\prime}}\frac{i}{\hbar}\left\langle V_{ll^{\prime}}\frac{\langle
l^{\prime}|\mathbf{\hat{r}|}l^{\prime\prime}\rangle V_{l^{\prime\prime}l}%
}{\epsilon_{l}-\epsilon_{l^{\prime}}-i\delta}\right\rangle _{c}%
+2\operatorname{Re}\sum_{l^{\prime},l^{\prime\prime}\neq l}\frac{-i}{\hbar
}\left\langle V_{ll^{\prime}}\frac{\langle l^{\prime}|\hat{V}|l^{\prime\prime
}\rangle\langle l^{\prime\prime}|\mathbf{\hat{r}|}l\rangle}{\epsilon
_{l}-\epsilon_{l^{\prime}}-i\delta}\right\rangle _{c}.
\end{equation}
Only the interband matrix elements (in the band-eigenstate basis) of the
position operator are relevant here. Utilizing $\langle\eta\mathbf{k}%
|\mathbf{\hat{r}|}\eta^{\prime}\mathbf{k}^{\prime}\rangle=i\frac{\partial
}{\partial\mathbf{k}}\delta_{\mathbf{kk}^{\prime}}\delta_{\eta\eta^{\prime}%
}+\mathbf{J}^{ll^{\prime}}$ with $\mathbf{J}^{ll^{\prime}}=\langle
u_{\mathbf{k}}^{\eta}|\frac{\partial}{\partial\mathbf{k}}|u_{\mathbf{k}}%
^{\eta^{\prime}}\rangle\delta_{\mathbf{kk}^{\prime}}$ and $\mathbf{J}%
^{l}\equiv\mathbf{J}^{ll}$, we get%
\begin{equation}
\delta^{ex}\mathbf{v}_{l}=\operatorname{Re}\sum_{l^{\prime}}\frac{2}{\hbar
}\left\langle V_{ll^{\prime}}\frac{\left[  V,\mathbf{J}\right]  _{l^{\prime}%
l}}{\epsilon_{l}-\epsilon_{l^{\prime}}-i\delta}\right\rangle _{c}%
+\sum_{l^{\prime}}\frac{2\pi}{\hbar}\left\langle \left\vert V_{ll^{\prime}%
}\right\vert ^{2}\right\rangle _{c}\delta\left(  \epsilon_{l}-\epsilon
_{l^{\prime}}\right)  \left[  i\mathbf{J}^{l^{\prime}}-i\mathbf{J}^{l}\right]
,
\end{equation}
where we define $\left[  V,\mathbf{J}\right]  _{l^{\prime}l}\equiv
\sum_{l^{\prime\prime}}\left[  V_{l^{\prime}l^{\prime\prime}}\mathbf{J}%
^{l^{\prime\prime}l}-\mathbf{J}^{l^{\prime}l^{\prime\prime}}V_{l^{\prime
\prime}l}\right]  $. This quantity can be greatly simplified by using
$\langle\eta\mathbf{k}|\mathbf{\hat{r}|}\eta^{\prime}\mathbf{k}^{\prime
}\rangle=i\frac{\partial}{\partial\mathbf{k}}\delta_{\mathbf{kk}^{\prime}%
}\delta_{\eta\eta^{\prime}}+\mathbf{J}^{ll^{\prime}}$:%
\begin{equation}
\left[  V,\mathbf{J}\right]  _{l^{\prime}l}=-i\langle l^{\prime}|\left[
\hat{V},\mathbf{\hat{r}}\right]  |l\rangle+\sum_{l^{\prime\prime}}\left[
\partial_{\mathbf{k}}\left(  V_{l^{\prime}l^{\prime\prime}}\delta
_{\mathbf{kk}^{\prime\prime}}\delta_{\eta\eta^{\prime\prime}}\right)
+\partial_{\mathbf{k}^{\prime}}\left(  \delta_{\mathbf{k}^{\prime}%
\mathbf{k}^{\prime\prime}}\delta_{\eta^{\prime}\eta^{\prime\prime}%
}V_{l^{\prime\prime}l}\right)  \right]  =\left(  \partial_{\mathbf{k}%
}+\partial_{\mathbf{k}^{\prime}}\right)  V_{l^{\prime}l},
\end{equation}
thereby%
\begin{equation}
\delta^{ex}\mathbf{v}_{l}=\sum_{l^{\prime}}\frac{2\pi}{\hbar}\left\langle
\left\vert V_{ll^{\prime}}\right\vert ^{2}\right\rangle _{c}\delta\left(
\epsilon_{l}-\epsilon_{l^{\prime}}\right)  \left[  i\mathbf{J}^{l^{\prime}%
}-i\mathbf{J}^{l}\right]  +\operatorname{Re}\sum_{l^{\prime}}\frac{2}{\hbar
}\left\langle \frac{V_{ll^{\prime}}\mathbf{\hat{D}}V_{l^{\prime}l}}%
{\epsilon_{l}-\epsilon_{l^{\prime}}-i\delta}\right\rangle _{c}%
\label{Luttinger}%
\end{equation}
with $\mathbf{\hat{D}}=\partial_{\mathbf{k}}+\partial_{\mathbf{k}^{\prime}}$.
This quantity is just the second term of Eq. (2.38) in Luttinger's classical
paper on the quantum transport theory of anomalous Hall effect
\cite{Luttinger1958}. In fact, the first term of Luttinger's Eq. (2.38) just
corresponds to the Berry-curvature anomalous velocity. Luttinger called his
Eq. (2.38) the off-diagonal velocity because its calculation involved
interband matrix elements of the velocity operator. This is the same thought
presented here. On the other hand, the second term of Luttinger's Eq. (2.38)
has been cited by Sinitsyn et al. \cite{Sinitsyn2006} to confirm the validity
of their pictorial definition of semiclassical side-jump velocity
$\mathbf{v}_{l}^{sj}=\sum_{l^{\prime}}\omega_{l^{\prime},l}^{2s}%
\delta\mathbf{r}_{l\prime,l}$ which contributes an anomalous Hall
current $\mathbf{j}^{sj}=\sum_{l}\mathbf{v}_{l}^{sj}g_{l}^{2s}$. The validity
of this definition of the side-jump velocity is finally
confirmed by the correspondence to Luttinger's quantum transport theory
\cite{Luttinger1958} and by one-to-one correspondence to special sets of
Feynman diagrams \cite{Sinitsyn2007,Kovalev2010,Nagaosa2010} as well as by
successful calculations of anomalous Hall effect in some model systems
\cite{Sinitsyn2007,Xiao2017AHE}. The last term of Eq. (\ref{Luttinger}) can be
split into two terms with one related to $\operatorname{Im}\left\langle
V_{ll^{\prime}}\mathbf{\hat{D}}V_{l^{\prime}l}\right\rangle _{c}$ and the
other related to $\mathbf{\hat{D}}\left\langle \left\vert V_{ll^{\prime}%
}\right\vert ^{2}\right\rangle _{c}$. The first one is related to the phase of
the disorder potential and is thus nontrivial. While the latter one that does
not break any symmetry is just a trivial renormalization
to $\mathbf{v}_{l}^{0}$. It does not contribute to the Hall current in the
leading order of perturbation theory and can be ignored
\cite{Sinitsyn,Sinitsyn2006}. In fact, in Rashba model (\ref{model}) with
short-range disorder and both bands partially occupied, this term vanishes.
Then one gets the relation $\delta^{ex}\mathbf{v}_{l}=$ $\mathbf{v}_{l}^{sj}%
$:
\[
\delta^{ex}\mathbf{v}_{l}=\sum_{l^{\prime}}\frac{2\pi}{\hbar}\left\langle
\left\vert V_{ll^{\prime}}\right\vert ^{2}\right\rangle _{c}\delta\left(
\epsilon_{l}-\epsilon_{l^{\prime}}\right)  \left[  i\mathbf{J}^{l^{\prime}%
}-i\mathbf{J}^{l}-\mathbf{\hat{D}}\arg V_{l^{\prime}l}\right]  \equiv
\sum_{l^{\prime}}\omega_{ll^{\prime}}^{2s}\delta\mathbf{r}_{l^{\prime}l}.
\]
\end{widetext}

As an example, considering the anomalous Hall effect in model (\ref{model})
with both bands partially occupied. By $\langle u_{\mathbf{k}}^{\eta
}|\mathbf{\hat{v}}|u_{\mathbf{k}}^{-\eta}\rangle=\frac{\alpha_{R}}{\hbar
}\mathbf{\hat{z}}\times\langle u_{\mathbf{k}}^{\eta}|\mathbf{\hat{\sigma}%
}|u_{\mathbf{k}}^{-\eta}\rangle$ and Eqs. (\ref{in}-\ref{inter-2}), we get
\begin{equation}
\delta^{in}\mathbf{v}_{l}=\frac{\alpha_{R}/\hbar}{\hbar/2}\mathbf{\hat{z}%
}\times\delta^{in}\mathbf{S}_{l},\text{ }\delta^{ex}\mathbf{v}_{l}%
=\frac{\alpha_{R}/\hbar}{\hbar/2}\mathbf{\hat{z}}\times\delta^{ex}%
\mathbf{S}_{l}.
\end{equation}
For class A impurities, one thus obtains zero anomalous Hall current under the
Gaussian disorder and%
\begin{equation}
\delta^{ex}\mathbf{v}_{l}=\eta\frac{\hbar\mathbf{k}_{\eta}\left(
\epsilon\right)  }{m}\times\mathbf{\hat{z}}\frac{J_{ex}\epsilon_{R}}{\left(
J_{ex}^{2}+2\epsilon_{R}\epsilon\right)  }\frac{\hbar}{2\Delta_{\eta}\left(
\epsilon\right)  \tau}.
\end{equation}
This result coincides with the side-jump velocity obtained in Ref.
\onlinecite{Xiao2017AHE} from the expression $\mathbf{v}_{l}^{sj}%
=\sum_{l\prime}\omega_{l\prime,l}\delta\mathbf{r}_{l\prime,l}$.

\section{Correspondence between semiclassical Boltzmann contributions and
Feynman diagrams in the band representation}

In the context of the anomalous Hall effect, the one-to-one correspondence
between semiclassical contributions and special sets of Feynman diagrams in
the band-eigenstate basis under the non-crossing approximation in the weak
disorder limit has been established \cite{Sinitsyn2007,Kovalev2010}. The
diagrams in the band representation for the disorder-induced
interband-coherence contributions to the anomalous Hall effect are presented
in Fig. 1 of Ref. \onlinecite{Kovalev2010}. The correspondence between these
diagrams and semiclassical contributions was clearly presented in Refs.
\onlinecite{Sinitsyn2007, Nagaosa2010}. The upper four \textquotedblleft
interband diagrams\textquotedblright\ with an interband velocity vertex on the
rhs of each diagram correspond to the semiclassical contribution due to the
anomalous distribution function $g_{l}^{a}$, the six \textquotedblleft
intraband diagrams\textquotedblright\ correspond to the semiclassical
contribution due to the intrinsic-skew-scattering $g_{l}^{sk-in}$. Whereas the
lower four \textquotedblleft interband diagrams\textquotedblright\ with an
interband velocity vertex on the left hand side of each diagram are just the
semiclassical contribution due to the side-jump velocity $\mathbf{v}_{l}^{sj}$.

In our case, this kind of correspondence remains unchanged, provided that the
left velocity vertex of all these diagrams in the case of the anomalous Hall
effect are replaced by the Feynman vertex of $A$. $\delta^{adis}A$ and
$\delta^{sk-in}A$, arising from $g_{l}^{a}$ and $g_{l}^{sk-in}$, are thus
represented by the upper four \textquotedblleft interband
diagrams\textquotedblright\ and the six \textquotedblleft intraband
diagrams\textquotedblright, respectively. As for $\delta^{sj}A$ which is
related to $\delta^{ex}A_{l}$, comparing the structure of Eqs. (\ref{inter-1}%
,\ref{inter-2}) with the left interband vertices and the interband-scattering
disorder lines in the lower four \textquotedblleft interband
diagrams\textquotedblright, one can verify the correspondence. This
correspondence is expected also because $\delta^{ex}A_{l}$ is a generalization
of the side-jump velocity.

The correspondence to the diagrammatic analysis establishes the equivalence
between the semiclassical theory and microscopic linear response theories
regarding disorder-induced interband-coherence responses under the
non-crossing approximation in the weak disorder limit. According to this
correspondence, it is clearly seen that $\delta^{adis}A$ and $\delta^{sj}A$
are both related to one interband and one intraband vertex, and are thus
interband-coherence disorder effects. While $\delta^{sk-in}A$\ is related to
two intraband vertices, it also contain interband-coherence disorder effects,
i.e., interband off-shell scattering processes, as shown by the middle part of
the six \textquotedblleft intraband diagrams\textquotedblright\ in Fig. 1 of
Ref. \onlinecite{Kovalev2010}. This point can be easily appreciated by
considering the case of 2D massive Dirac model \cite{Sinitsyn2007,Kohno2014},
where the interband impurity-scattering can only be virtual transition.

\section{SOT in a 2D Rashba ferromagnet with in-plane magnetization and scalar
impurities}

The model Hamiltonian is \cite{Zhang2008} $H=\frac{\hbar^{2}\mathbf{k}^{2}%
}{2m}+\alpha_{R}\mathbf{\hat{\sigma}}\cdot\left(  \mathbf{k}\times
\mathbf{\hat{z}}\right)  -J_{ex}\mathbf{\hat{\sigma}}\cdot\mathbf{\hat{M}%
}+V_{A}$, with $\mathbf{\hat{M}=}\cos\theta_{M}\mathbf{\hat{x}}+\sin\theta
_{M}\mathbf{\hat{y}}$ the direction of the in-plane magnetization. The
eigenenergy of the pure system is $\epsilon_{\mathbf{k}}^{\eta}=\frac
{\hbar^{2}k^{2}}{2m}+\eta\Delta_{\mathbf{k}}$, where $\Delta_{\mathbf{k}%
}=\left\vert \alpha_{R}\left(  \mathbf{k}\times\mathbf{\hat{z}}\right)
-J_{ex}\mathbf{\hat{\sigma}}\cdot\mathbf{M}\right\vert $. Note that $k\equiv
k\left(  \phi\right)  $ still depends on $\phi$ due to the anisotropy of the
bands arising from the interplay of Rashba effective magnetic field and
in-plane magnetization. The spinor eigenstate reads $|u_{\eta\mathbf{k}%
}\rangle=\frac{1}{\sqrt{2}}\left[  1,-i\eta\exp\left(  i\gamma_{\mathbf{k}%
}\right)  \right]  ^{T}$ with $\cos\gamma_{\mathbf{k}}=\frac{J_{ex}\sin
\theta_{M}+\alpha k_{x}}{\Delta_{\mathbf{k}}}$ and $\sin\gamma_{\mathbf{k}%
}=\frac{-J_{ex}\cos\theta_{M}+\alpha k_{y}}{\Delta_{\mathbf{k}}}$. In order to
make analytical progress, we focus on the limit \cite{Zhang2008} $\hbar
/\tau\ll\alpha k_{F}\ll J_{ex}\ll\epsilon_{F}$. The following expressions are
obtained by expanding to the first order of $\alpha_{R}k/J_{ex}$:
\begin{align*}
\Delta_{\mathbf{k}}  &  \simeq J_{ex}\left[  1+\frac{\alpha_{R}k}{J_{ex}}%
\sin\left(  \theta_{M}-\phi\right)  \right]  ,\\
\cos\gamma_{\mathbf{k}}  &  \simeq\sin\theta_{M}+\frac{\alpha_{R}k}{J_{ex}%
}\left[  \cos\phi-\sin\theta_{M}\sin\left(  \theta_{M}-\phi\right)  \right]
,\\
\sin\gamma_{\mathbf{k}}  &  \simeq-\cos\theta_{M}+\frac{\alpha_{R}k}{J_{ex}%
}\left[  \sin\phi+\cos\theta_{M}\sin\left(  \theta_{M}-\phi\right)  \right]  ,
\end{align*}
and $\sin\left(  \gamma_{\mathbf{k}^{\prime}}-\gamma_{\mathbf{k}}\right)
\simeq\frac{\alpha k^{\prime}}{J_{ex}}\cos\left(  \theta_{M}-\phi^{\prime
}\right)  -\frac{\alpha k}{J_{ex}}\cos\left(  \theta_{M}-\phi\right)  $,
$\cos\left(  \gamma_{\mathbf{k}^{\prime}}-\gamma_{\mathbf{k}}\right)  \simeq1$.

Under a weak uniform electric field applied in x direction, the intrinsic
nonequilibrium spin-polarization is given by $\delta^{in}\mathbf{S}\simeq
\frac{\hbar}{2}eD_{0}\frac{\alpha_{R}\cos\theta_{M}}{J_{ex}}E_{x}%
\mathbf{\hat{z}}$. As for $\delta^{ex}\mathbf{S}_{l}$, in the weak scattering
limit the nonzero component in the first order of $\alpha_{R}k/J_{ex}$ is
$\delta^{ex}\mathbf{S}_{l}\simeq\frac{\hbar}{2}\eta\frac{\hbar}{2J_{ex}\tau
}\frac{\alpha_{R}k}{J_{ex}}\cos\left(  \theta_{M}-\phi\right)  \mathbf{\hat
{z}}$. Thus in the $o\left(  \alpha_{R}k/J_{ex}\right)  $ contribution of
$\delta^{sj}\mathbf{S}=\sum_{l}\delta^{ex}\mathbf{S}_{l}g_{l}^{2s}$ the
distribution function can be obtained from the Boltzmann equation in the
zeroth order of $\alpha_{R}k/J_{ex}$, just yielding $g_{\eta}^{2s}\left(
\epsilon,\phi\right)  =e\mathbf{E}\cdot\frac{\hbar\mathbf{k}_{\eta}}{m}%
\tau\left(  -\partial_{\epsilon}f^{0}\right)  $. With only in-plane
magnetization there is no anomalous distribution function and intrinsic skew
scattering. Then the scattering-induced interband-coherence contribution is
obtained as $\delta^{SJ}\mathbf{S}=\delta^{sj}\mathbf{S}=-\delta
^{in}\mathbf{S}$, which cancels the intrinsic nonequilibrium
spin-polarization. This vanishing interband-coherence contribution to the
nonequilibrium spin-polarization is consistent with the result in Ref.
\onlinecite{Titov2017}.
%Similarly, under an electric field applied in the y direction one has the same cancellation.


\begin{thebibliography}{99}
\bibitem {note}More discussions about this point are given in Sec. IV. A.
\bibitem {note-SOscattering}In this paper we do not consider model
Hamiltonians with spin-orbit scattering.
\end{thebibliography}
\end{document}